\documentclass{article}

\usepackage{arxiv}
\usepackage{cite}
\usepackage[utf8]{inputenc} % allow utf-8 input
\usepackage[T1]{fontenc}    % use 8-bit T1 fonts
\usepackage{hyperref}       % hyperlinks
\usepackage{url}  
\usepackage{graphicx}
\usepackage{float}
\usepackage{booktabs}       % professional-quality tables
\usepackage{amsfonts}       % blackboard math symbols
\usepackage{nicefrac}       % compact symbols for 1/2, etc.
\usepackage{microtype}      % microtypography
\usepackage{lipsum}
\title{Randgan: randomized generative adversarial network for detection of covid-19 in chest X-ray}

\author{
  Saman Motamed \\
  \texttt{sam.motamed@mail.utoronto.ca} \\
   \And
  Patrik Rogalla \\
  \texttt{patrik.rogalla@uhn.ca} \\
  \And
 Farzad Khalvati \\
  \texttt{farzad.khalvati@utoronto.ca} \\
  \\
  Intelligent Medical Image Computing Systems (IMICS) Lab\\
  \url{www.imics.ca}\\
  Department of Medical Imaging\\
  The Hospital for Sick Children\\
  University of Toronto \\
  Toronto, Ontario, Canada\\
}
\begin{document}
\maketitle

\begin{abstract}
COVID-19 spread across the globe at an immense rate has left healthcare systems incapacitated to diagnose and test patients at the needed rate. Studies have shown promising results for detection of COVID-19 from viral bacterial pneumonia in chest X-rays. Automation of COVID-19 testing using medical images can speed up the testing process of patients where health care systems lack sufficient numbers of the reverse-transcription polymerase chain reaction (RT-PCR) tests. Supervised deep learning models such as convolutional neural networks (CNN) need enough labeled data for all classes to correctly learn the task of detection. Gathering labeled data is a cumbersome task and requires time and resources which could further strain health care systems and radiologists at the early stages of a pandemic such as COVID-19. In this study, we propose a randomized generative adversarial network (RANDGAN) that detects images of an unknown class (COVID-19) from known and labelled classes (Normal and Viral Pneumonia) without the need for labels and training data from the unknown class of images (COVID-19). We used the largest publicly available COVID-19 chest X-ray dataset, COVIDx, which is comprised of Normal, Pneumonia, and COVID-19 images from multiple public databases. In this work, we use transfer learning to segment the lungs in the COVIDx dataset. Next, we show why segmentation of the region of interest (lungs) is vital to correctly learn the task of classification, specifically in datasets that contain images from different resources as it is the case for the COVIDx dataset. Finally, we show improved results in detection of COVID-19 cases using our generative model (RANDGAN) compared to conventional generative adversarial networks (GANs) for anomaly detection in medical images, improving the area under the ROC curve from 0.71 to 0.77. 
\end{abstract}

% keywords can be removed

\section{Introduction}
COVID-19 spread globally over a short period of time and became a deadly pandemic~\cite{du2020predictors}. Early diagnosis and detection of pneumonia can minimize the risk factors of the illness~\cite{pereira2012severe} and help break the transmission chain. The standard test for diagnosis of COVID-19 is reverse transcriptase polymerase chain reaction (RT-PCR)~\cite{wang2020detection}. The lack of accessibility and slowness of RT-PCR, along with its high false negative rate \(39 - 61\% \), drew attention to diagnosis of COVID-19 using chest radiographs~\cite{kundu2020might, kucirka2020variation}. Automation of COVID-19 diagnosis using chest X-rays can help healthcare systems keep up with demands for patients testing as X-rays are more readily available than RT-PCR and reduce strain from radiologists and healthcare systems. Medical imaging based diagnosis can also help control the high false negative rate of RT-PCR tests by acting as a secondary control. Computer-aided disease diagnosis using medical imaging techniques have accelerated over the past decade due to the breakthroughs in the field of Machine Learning and the development of detection and classification models that are based on Convolutional Neural Networks (CNNs)~\cite{Long_2015_CVPR , girshick2015fast, krizhevsky2012imagenet}. CNNs, which are mainly used in supervised frameworks, require large amounts of labeled data to learn the task of anomaly detection, such as detecting COVID-19 in chest X-rays. Supervised architectures require training data with complete labels for all image classes (e.g., normal and COVID-19). Nevertheless, this requires accurate labeling of the data for all cases and the cumbersome annotation effort, and the diagnosis variation amongst expert radiologists limits the performance of these supervised models on new data. Specially, in pandemics such as COVID-19, at the beginning, there is limited COVID-19 data (if any data at all) available for training a supervised classification model. In contrast, solutions based on semi-supervised learning only require partial labels for the training data \cite{zhu2009introduction}. Semi-supervised learning significantly reduces the cost of creating training data and ,thus, opens new opportunities for automated disease detection using training data with only single class labels.

\par In this study, we propose a semi-unsupervised generative model (RANDGAN) for detection of COVID-19 positive chest X-ray images. The idea behind anomaly detection using generative adversarial networks (GANs) comes from the great ability of generative models in learning the image-space manifold where training images lie on, and being able to generate never-before-seen images that lie on the learned image-space~\cite{goodfellow2014generative}. Anomaly detection may be seen as only detecting abnormality in medical images such as a tumour or pneumonia. We extend the definition of anomaly in medical images as the deviation from the image-space manifold of training data. In other words, if the training data only includes COVID-19 negative cases (i.e., healthy or viral pneumonia), the anomaly detected in test cases is indeed an abnormality such as COVID-19. On the other hand, if the training data only includes COVID-19 positive cases, the "anomaly" detected in the test cases are the deviation from COVID-19 cases, meaning that the test case does not contain the abnormality in the training class (i.e., healthy or viral pneumonia). We show our proposed RANDGAN model is able to differentiate between COVID-19 positive and negative images. To the best of our knowledge, this study is the first of its kind, using semi-supervised learning for detection of COVID-19 in medical images and reporting performance accuracy on the entire cohort of COVID-19 positive images without the need to use any of the COVID-19 positive images to train our model.

\section{Dataset}
\begin{figure}[h!]
  \centering
  \includegraphics[scale=0.6]{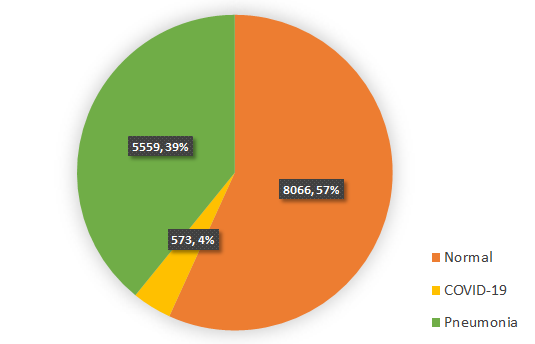}
  \caption{Class distribution of COVIDx dataset}
  \label{fig:1}
\end{figure}

Covid-chestxray dataset~\cite{cohen2020covid} is an effort by Cohen~\emph{et al.} to make a public COVID-19 dataset of chest X-ray images with normal, pneumonia, and COVID-19 radiological readings. Wang~\emph{et al.} uses covid-chestxray dataset,  along with four other publicly available datasets and compiles COVIDx~\cite{wang2020covid} dataset. With the number of images growing, many deep learning models are trained and tested on this public dataset~\cite{wang2020covid, ozturk2020automated, afshar2020covid}. Figure~\ref{fig:1} shows the class distribution of the COVIDx dataset. The images are in RGB format, with pixel range of \([0 , 255]\) and have various sizes. To train the generative models in this study, all images were converted to gray scale, resized to $128 \times 128$ pixels and normalized to have pixel intensities in the \([-1 , 1]\) range.

\section{Related Work}
Using the covid-chestxray and COVIDx datasets, multiple studies have utilized supervised deep learning models to detect COVID-19 in chest X-rays~\cite{wang2020covid, ozturk2020automated, karim2020deepcovidexplainer, hemdan2020covidx, ghoshal2020estimating, afshar2020covid}. Wang~\emph{et al.}'s CNN based COVID-NET~\cite{wang2020covid} achieved a 93.3\% test accuracy for multi-class classification on a test cohort of 100 Normal, 100 Pneumonia and 100 COVID-19 from the COVIDx dataset with the rest of images of each class being used to train their model. Hemdan~\emph{et al.}'s COVIDX-Net~\cite{hemdan2020covidx}, comprised of multiple architectures such as VGG19, DenseNet121 and InceptionV3, was tested on a small set of 50 X-ray images from the covid-chestxray dataset. 25 COVID-19 positive and 25 COVID-19 negative. They report accuracy of anywhere between 50\% (InceptionV3) to 90\% (VGG19 and DenseNet201) for each investigated architecture. Ozturk \emph{et al.}'s DarkNet~\cite{ozturk2020automated} experimented with both binary classification (COVID-19 vs. No Findings) and multi-class classification (Pneumonia vs. COVID-19 vs. No Findings). They report a binary classification accuracy of 98.08 \% and multi-class classification with accuracy of 0.87\% on 25 COVID-19, 100 Normal and 100 Pneumonia images.
Afshar~\emph{et al.} proposed using capsule networks for binary classification of COVID-19 positive and negative cases using COVIDx dataset, pre-trained on non-COVID chest X-ray images from other datasets. They report an Accuracy of \(95.7\%\), Sensitivity of \(90\%\), Specificity of \(95.8\%\), and the area under the ROC curve (AUC) of 0.97. The number of test images from each class is not disclosed in their paper.

\par The high accuracy achieved in these models, despite the imbalanced dataset with only \(4\%\) of the images belonging to COVID-19 and the multi-centric nature of the dataset which could cause images from different scanners and health centres to have inherent difference in characteristics, put the robustness of these models under question. Another issue is transparency in number of test images used in these studies where the train and test split of the dataset used in the experiments is not clear. 

DeGrave~\emph{et al.} \cite{degrave2020ai} conducted a few experiments to test the generalizability and robustness of these models trained on the COVIDx dataset. Replicating different supervised models such as COVID-NET~\cite{wang2020covid} and training the models on COVIDx dataset, they achieve high test accuracy when tested on COVIDx data. Their predictive performance, however, drops by 50\% when they validate their model on an external COVID and Non-COVID dataset~\cite{vaya2020bimcv} where the images are from a single institute. Furthermore, using saliency maps, which highlight the region of each X-ray image that contributed most to the classification decision of the CNNs, they find non-COVID markers such as image edges, diaphragm and cardiac silhouette have contributed to the classification of COVID-19; markers that do not have a predictive value for detection of COVID-19~\cite{ng2020imaging}. This confirms that using the full images from a dataset that comes from different scanners can be problematic where non-disease specific markers could act as a shortcut~\cite{geirhos2020shortcut} and help CNNs achieve high accuracy on a particular dataset yet fail to generalize to any other dataset. To minimize the effect of shortcuts, we create a segmented COVIDx dataset that includes only the lungs where the true markers of COVID-19 and Pneumonia appear.
\section{Segmentation of Lung in COVIDx Images}
To mitigate the issue of deep learning models picking non-disease related markers from the images, we created a new dataset by segmenting the lungs of COVIDx dataset. As the training set, we used the Montgomery County chest X-ray set~\cite{jaeger2014two}, which contains 138 frontal chest X-rays from Montgomery County’s Tuberculosis screening program with corresponding masks manually annotated radiologists. We resized the images to \(256 \times 256\) pixels and normalized to have pixel intensities between 0 and 1. We trained a U-NET~\cite{Ronneberger_2015} based model, that has been augmented with Inception and Residual architectures, with these normalized images~\cite{motamed2019transfer, clark2017fully}. Transfer learning~\cite{tan2018survey} has shown promise in adapting tasks from one domain (source) to another (target). 
For the task of lung segmentation for COVIDx dataset, the Montgomery dataset was used as the source and COVIDx as the target domains. For the task of transfer learning, Sefexa~\cite{sefexa}, an image segmentation tool, was used for manual segmentation of 900 randomly selected images of the COVIDx dataset. All segmentation masks were corrected by an experienced radiologist and intentionally over-segmented to ensure no region of lung is excluded. Thus, these masks are best to be used for classification algorithms for detection of COVID-19 and pneumonia, and not for precise segmentation of lung boundaries. 850 segmented X-ray images from COVIDx were used for performing transfer learning~\cite{motamed2019transfer} from the Montgomery dataset to COVIDx. We kept 50 manual segmentations to evaluate our model's accuracy. Since source domain is smaller than our target domain, we fine-tuned 75\% of the pre-trained model’s layers (encoder part), and trained on the Montgomery dataset images. We froze the first 25\% layers of the pre-trained U-NET and fine-tuned the rest of the encoder and decoder components based on our manual segmentation for COVIDX images. We used open and close operations as a post-processing step to fill any holes in the masks and reduce noise in the predicted masks. We tested the accuracy of our model using Sørensen–Dice coefficient (DSC). We achieved a DSC of $0.83$ on our test set of 50 images. 

\begin{figure}[h!]
  \centering
  \includegraphics[scale=0.5]{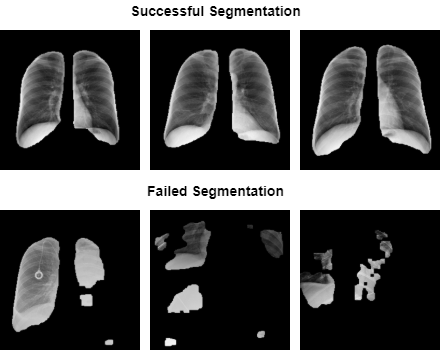}
  \caption{Output samples of our segmentation model on COVIDx images}
  \label{fig:2}
\end{figure}

Figure~\ref{fig:2} shows the output of our segmentation model. We include some of the failed segmentation attempts of the model as well. Accurate segmentation of lung images are a limitation of using automated segmentation models.

\section{Random Input Generative Adversarial Networks}

Generative Adversarial Networks (GANs)~\cite{goodfellow2014generative} revolutionized the field of deep learning by allowing generation of never-before-seen data that follows the distribution of real data. Applications of GANs have expanded from generating human-like faces, to image style transfer and detection of anomalies in images \cite{radford2015unsupervised}. Below we describe the components of our proposed Random Input Generative Adversarial Networks (RANDGAN).

\subsection{Generator Network}
\begin{figure}
  \centering
  \includegraphics[scale=0.4]{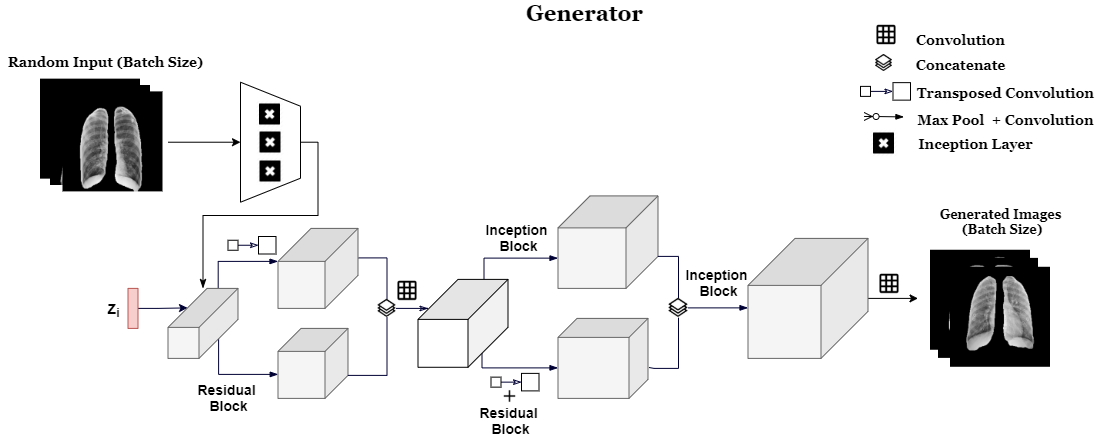}
  \caption{RANDGAN's Generator Architecture}
  \label{fig:3}
\end{figure}

\begin{figure}
  \centering
  \includegraphics[scale=0.5]{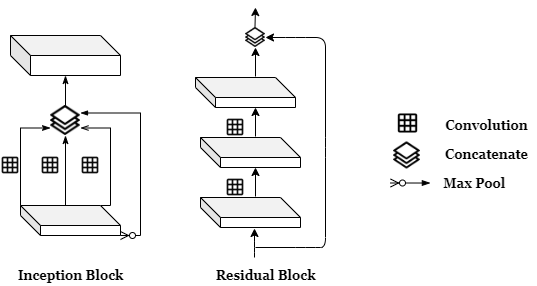}
  \caption{Inception and Residual Block Architecture}
  \label{fig:4}
\end{figure}

The Generator (\textbf{G}) (Fig.~\ref{fig:3}) learns a distribution $P_g$ over the input data $x$ via mapping of input noise $z$, to $2D$ images by function $G(z)$.
The trained Generator learns the mapping $G(z): z \longmapsto x$ from latent space representations $z$ to realistic, $2D$, X-ray images. Our Generator model follows DCGAN's architecture (named AnoGAN for anomaly detection GAN in the study) ~\cite{radford2015unsupervised} (used for anomaly detection for retina) with three main modifications; the use of randomized 2D image inputs to the generator, inception layers, and  residual connections as shown in Fig.~\ref{fig:4}.

Feeding real training image as an input to the generator has shown improvement in using GANs for augmenting images \cite{antoniou2017data, motamed2020inception}. Real image is encoded into a lower dimensional space before being concatenated with the noise input vector \(z\). To improve generalizability of our generator, specially due to the multi-centric nature of COVIDx data, we randomly select \(batch-size\) images from the cohort of our training class and encode them to a lower-representation space using inception layers. This helps in adding variability to each iteration of the generator's training by not only using a random noise vector, but also real, random image representations of the training class. Doing so shows improved results when using the trained GAN to classify images of class c from other classes.
The idea behind the inception and residual architecture~\cite{szegedy2016rethinking} is being able to increase GAN's ability to capture more details from training image-space without losing spatial information after each convolution and pooling layer. Although making the Generator deeper is theoretically a valid way to capture more long-range details in the image, deep GANs are unstable and hard to train~\cite{radford2015unsupervised, kodali2017convergence}.  

\subsection{Discriminator}
\begin{figure}
  \centering
  \includegraphics[scale=0.45]{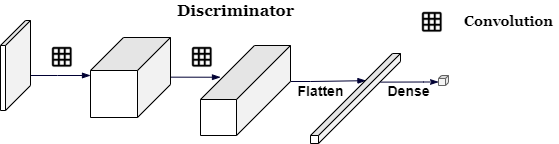}
  \caption{RANDGAN's Discriminator Architecture}
  \label{fig:5}
\end{figure}

The Discriminator (\textbf{D}) (Fig.~\ref{fig:5}) is a 4-layer CNN that maps a 2D image to a scalar output that can be interpreted as the probability of the given input being a real chest X-ray sampled from training data or generated G(z) by the Generator G. 

Optimization of D and G can be thought of as the following game of minimax~\cite{goodfellow2014generative} with the value function $V(G, D)$: 
\begin{equation} \label{eq:1}
\min_G \max_D V(D, G) = \mathbb{E}_{x_{{\sim_P}_{data{(x)}}}} [\log D(x)] + \mathbb{E}_{z_{{\sim_P}_{z{(z)}}}} [\log (1 - D(G(z)))]
\end{equation}

During training, Generator G is trained to minimize the accuracy of Discriminator D's ability in distinguishing between real and generated images while the Discriminator is trying to maximize the probability of assigning real training images the ``real" and generated images from G, ``fake" labels. The Generator improves at generating more realistic images while Discriminator gets better at correctly identifying between real and generated images.

\section{Experiments}
\subsection{Data and Pre-processing}
We used both full images from COVIDx dataset and our segmentation of the COVIDx data to train separate models and compare the results. One of the advantages of our semi-supervised model compared to supervised models is the ability to test our model on not only a subset, but all of COVID-19 positive images as we do not use any of the images to train our model. While studies such as Wang et. al's COVID-NET use 100 images of COVID-19, Hedman~\emph{et al.}\cite{hemdan2020covidx} and Ozturk~\emph{et al.}\cite{ozturk2020automated} using 25 COVID-19 positive images to test their models, we used 573 images of each class; the entire dataset for COVID-19 was used and for normal and pneumonia classes, 573 images were randomly selected for each class. All images, converted from RGB to grayscale, were resized to 128 $\times$ 128 pixels, with pixel intensities normalized to have values between -1 and 1. tHe models were trained using an NVIDIA GeForce RTX 2080 Ti with 11 GB of memory.

Table \ref{table:1} shows the Train and Test split of our COVIDx and Segmented COVIDx images.

\begin{table}[h!]
\begin{center}
\caption{Train and Test class distribution of COVIDx and COVIDx segmentation dataset}
\label{table:1}
\begin{tabular}{ | c | c | c | }
  \hline
  Label & Train & Test \\ \hline
  Normal & 7,493 & 573 \\ \hline
  Pneumonia & 4,986 & 573\\ \hline
  COVID-19 & N/A & 573 \\ \hline
\end{tabular}
\end{center}
\end{table}

\subsection{Evaluation}
    We trained two instances of our RANDGAN. For comparison, we repeated the same training using the GAN model (AnoGAN) used in Radford et. al's~\cite{radford2015unsupervised} anomaly detection study. One RANDGAN / AnoGAN was trained using Normal images and the other RANDGAN / AnoGAN was trained using Pneumonia images. When the model's training is done, the Generator has learned the mapping $G(z): z \longmapsto x$ from latent space representation $z$ to realistic images (Chest X-ray with Pneumonia). Given a query image $x$ in test, we want to find a point $z$ from the latent space that, given the Generator's output on that point ($G(z)$ ), that is most similar to the query image $x$. The expected behaviour after successful training is that the query image $x$, if affected by pneumonia, will result in finding an image $G(z)$, which is visually closer to image $x$ than if the query image was a normal case.
\par To find latent variable $z$ that generates the most similar image $G(z)$ to the query image $x$, we used back propagation with a predefined number of steps. The loss function defined to find such $z$ through back-propagation is comprised of two components; \textit{residual loss} and \textit{discrimination loss}. Residual loss ($\mathcal{L_R}$) calculates the L1 distance between $G(z)$ and the query image $x$ and enforces visual similarity between the query image and generated image. 
\begin{equation}
\mathcal{L}_R({z_i}) = \sum|x - G(z_i)|
\end{equation}

\par Schlegl \emph{et al.}~\cite{schlegl2017unsupervised} proposed a discrimination loss ($\mathcal{L_D}$) inspired by the concept of feature matching~\cite{salimans2sama016improved} that enforces generated images $G(z_i)$ to follow the statistical characteristics of the training images. $\mathcal{L_R}$ is defined below where the output of an intermediate layer of the discriminator, $f(.)$, is used to represent the statistical characteristics of the input image. 
\begin{equation}
\mathcal{L}_D({z_i}) = \sum|f(x) - f(G(z_i))|
\end{equation}

\par The overall loss used to back-propagate and find the best z is a weighted sum of residual and discrimination loss; 
\begin{equation}
    \mathcal{L}({z_i}) = (1 - \lambda) \times \mathcal{L}_{R}({z_i}) + \lambda \times \mathcal{L}_{D}({z_i})
\end{equation}

The Anomaly score $A(x)$ for the query image $x$ is defined as;
\begin{equation} \label{eq:ax}
A(x) = (1 - \lambda) \times R(x) + \lambda \times D(x)
\end{equation}

\noindent where R(x) and D(x) are respectively the residual and discrimination loss of the best $z_i$ found through back-propagation. $\lambda$ adjusts the weighted sum of the overall loss and anomaly score. We used $\lambda = 0.2$ to train our proposed RANDGAN and AnoGAN~\cite{schlegl2017unsupervised}. Both architectures were trained with the same initial conditions for performance comparison.
\par With two trained models, one on Normal and one on Pneumonia images, we calculate two anomaly scores \ref{eq:ax} for each test image. One anomaly score from inputting the test image into Normal trained GAN and one from Pneumonia trained GAN. The anomaly score generated from the Normal trained GAN will be lower for Normal test images compared to Pneumonia and COVID-19 images. Respectively, the anomaly score generated from Pneumonia trained GAN will be lower for Pneumonia test images compared to Normal and COVID-19 images. For each test image and the corresponding two anomaly scores, we generate a single anomaly score by summing the two scores together. The idea is that COVID-19 (unknown) images would score high anomalies from both networks while Normal and Pneumonia images score low in one model and high in the other. This should lead to the COVID-19 (unknown) images to score higher overall than the two other (known) classes.
\section{Results}
We generated a single anomaly score, comprised of two anomaly scores from the two trained models (Normal, Pneumonia), for the images in our test set. 573 anomaly scores were computed for each class (Normal, Pneumonia and COVID-19) of our COVIDx and segmented COVIDx dataset. To evaluate the performance of our COVID-19 positive detection model on a balanced test set, we randomly selected 286 Normal labeled and 287 Pneumonia labeled images and combine them into a COVID-19 negative test set with corresponding anomaly scores. We repeated the random selection of images from Normal and Pneumonia test cohorts 5 times in order to achieve an average performance metric of our models. The experiments were performed using AnoGAN trained on full COVIDx images, AnoGAN trained on segmented COVIDx images and RANDGAN on segmented COVIDx images. Table \ref{table:2} shows the average AUC of our models for the 5 calculations. We also report the AUC on the unbalanced test set, using 573 COVID-19 positive and 1146 COVID-19 negative (573 normal and 573 Pneumonia) images.

\begin{table}[h!]
\begin{center}
\caption{Performance comparison of RANDGAN and AnoGAN}
\label{table:2}
\begin{tabular}{ | c | c | c | }
  \hline
  Model & Dataset & AUC \\ \hline
  AnoGAN & COVIDx (Balanced test set)& 0.54  \\ \hline
  AnoGAN & Segmented COVIDx (Balanced test set) & 0.71 \\ \hline
  RANDGAN & Segmented COVIDx (Balanced test set) & \textbf{0.77} \\ \hline
  RANDGAN & Segmented COVIDx (Imbalanced test set) & 0.76  \\ \hline
\end{tabular}
\end{center}
\end{table}

\begin{figure}[h!]
  \centering
  \includegraphics[scale=0.35]{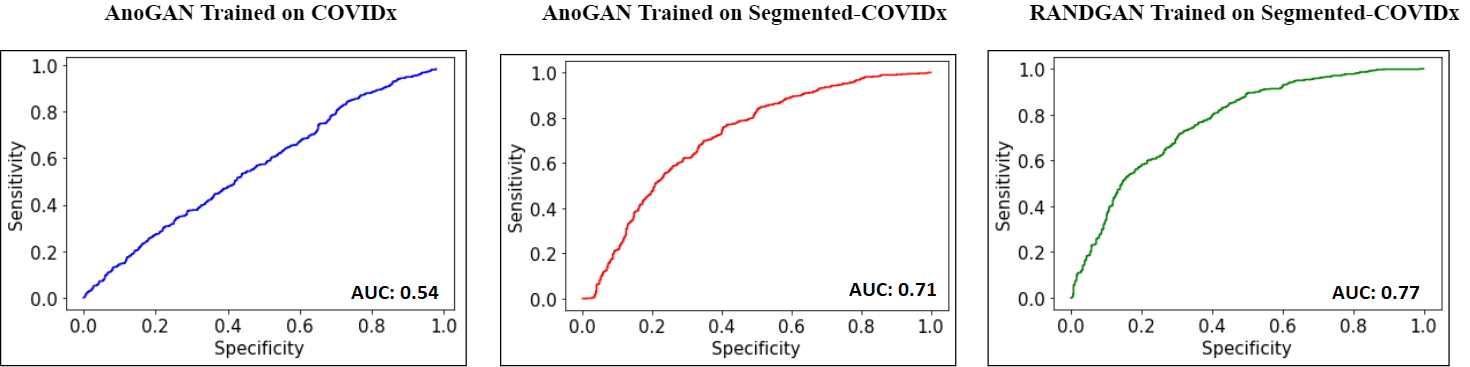}
  \caption{ROC curve of the trained generative models}
  \label{fig:roc}
\end{figure}

Figure \ref{fig:roc} shows the Receiver operating characteristic (\textbf{ROC}) curve of the 3 trained models.

\begin{figure}
  \centering
  \includegraphics[scale=0.5]{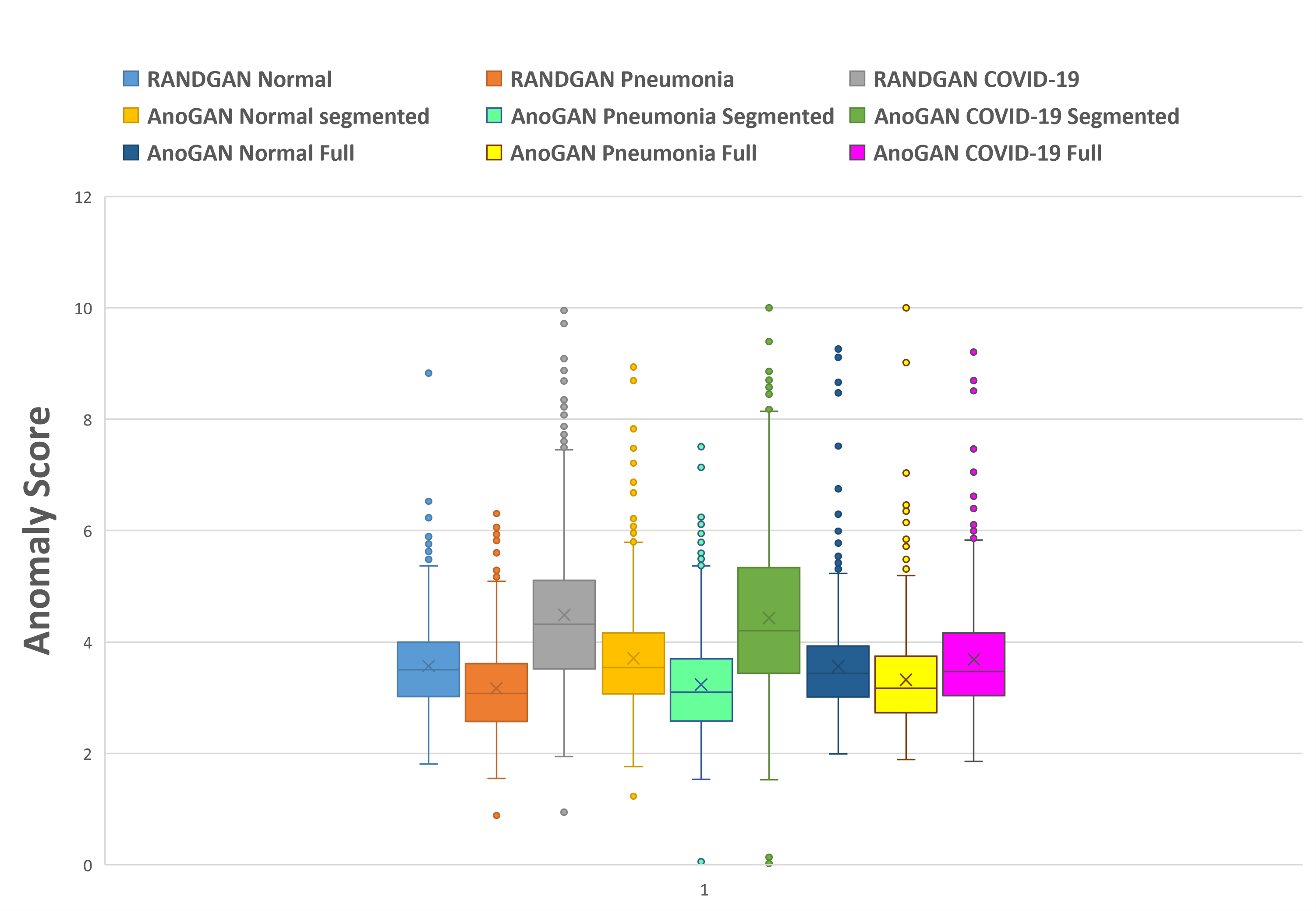}
  \caption{Normalized average anomaly score of the trained generative models}
  \label{fig:ano}
\end{figure}

Figure \ref{fig:ano} shows the normalized average anomaly score of the 5 runs of each of our three models; RANDGAN trained on segmented X-ray images, AnoGAN trained on segmented images and AnoGAN trained on full images. Despite the ROC curve that combines a balanced number of Normal and Pneumonia images in comparison to COVID-19 images, we present the anomaly scores in their entirety (573 Normal, 573 Pneumonia and 573 COVID-19 images). RANDGAN shows the biggest gap of normalized mean anomaly score (MAS) between COVID-19 (MAS = 4.48) and Pneumonia (3.01) and COVID-19 and Normal (3.56) images which are 1.47 and 0.92 respectively. AnoGAN trained on segmented COVIDx dataset shows 1.36 as the gap between COVID-19 (MAS = 4.42) and 0.91 between COVID-19 and Normal (3.51). AnoGAN trained on full COVIDx images shows a small gap between COVID-19, Pneumonia and Normal images (0.36 between COVID-19 and Pneumonia and 0.12 between COVID-19 and Normal).
\section{Discussion}
In this study, we introduced RANDGAN, a novel generative adversarial network for semi-supervised detection of an unknown (COVID-19) class in chest X-ray images from a pool of known (Normal and Pneumonia) and unknown classes (COVID-19) by only using the known classes for training. With this model, unknown cases can be screened and flagged for further investigations by radiologists increasing the probability of catching such cases early on. Using semi-supervised approaches for a problem such as detection of COVID-19, specially at the beginning of a pandemic are preferred over supervised approaches for they allow faster training of models without the need for gathering and annotation of data from the spreading disease. The result of semi-supervised models are also more robust where number of images are limited for the unknown (COVID-19) class because all images can be used to test the model where supervised models have to use majority of the images for training the model and test the model on a small subset of the images.

\par We also showed the importance of segmentation of lungs for the COVIDx dataset. DeGrave \emph{et al.} \cite{degrave2020ai} showed non-disease markers outside the lung act as shortcuts \cite{geirhos2020shortcut} in helping CNNs performance on specific datasets on which, the model is trained. By using transfer learning and segmenting the lung, we showed that using lung only images boosts the performance of generative models in detecting COVID-19 from Pneumonia and Normal images. AnoGAN~\cite{radford2015unsupervised} achieved an average AUC of 0.54 when using full images from COVIDx images while using segmented COVIDx images achieved an average AUC of 0.71. 

Future directions will focus on improving the performance of our proposed RANDGAN (AUC of 0.77) model by performing data augmentation and as more data is collected, it is important to validate the model on external data sources.

\section{Limitations}
One limitation of working with data of relatively early stages of a disease such as COVID-19 is dataset size. Even though our semi-supervised model is able to use all COVID-19 images to evaluate the performance of the model, while supervised models have to use majority of the already small COVID-19 cohort to train their images, more images would allow for a better understanding of the true performance of both supervised and semi-supervised models. Segmentation accuracy of the lungs is another limiting factor. Although the performance of the base model greatly improves (AUC of 0.54 to 0.71), segmentation model fails in some cases (Figure ~\ref{fig:2}). As more data gets collected and becomes available from different health care systems, any model trained for detection of COVID-19 needs validation from external sources. Without validation, these models need to be used as a secondary measure for detection of COVID-19.

\bibliography{references}
\end{document}